\documentclass[11pt,a4paper]{article}

\pdfoutput=1
\usepackage{amsmath,amsfonts,amsbsy,amssymb,enumerate,array}
\usepackage[usenames]{color}

\definecolor{AV}{rgb}{0.65,0.0,0}
\definecolor{DT}{rgb}{0,0,0.65}

\newcommand{\rom}[1]{\mathrm{#1}}
\definecolor{purple}{rgb}{0.7,0,1}
\definecolor{grey}{rgb}{0.5,0.5,0.5}

\def\ws{{w^{I_1}}}
\def\bs{{b^{I_1}}}
\newcommand{\ben}{\begin{eqnarray}\displaystyle}
\newcommand{\een}{\end{eqnarray}}

\usepackage[colorlinks=true,      linkcolor=blue,      urlcolor=blue,      filecolor=blue,      citecolor=blue,
      pdfstartview=FitH,     pdfpagemode=UseNone,      bookmarksopen=true]{hyperref}  

\usepackage[all]{hypcap}     

\numberwithin{equation}{section}

\usepackage{braket}

\newcommand{\del}{\partial}
\newcommand{\invert}[1]{\frac{1}{#1}}

\newcommand{\nn}{\nonumber}
\newcommand{\be}{\begin{equation}}
\newcommand{\ee}{\end{equation}}
\newcommand{\beeq}{\begin{eqnarray}}
\newcommand{\eeeq}{\end{eqnarray}}
\newcommand{\bea}[1]{\begin{align}#1\end{align}}
\newcommand{\grad}{{\rm {\bf \nabla}}}

\newcommand{\e}{\epsilon}
\newcommand{\m}{\mu}							
\newcommand{\n}{\nu}								
\renewcommand{\l}{\lambda}								
\newcommand{\p}{\psi}								
\newcommand{\q}{\theta}		
\renewcommand{\r}{\rho}		

\newcommand{\lb}{\left (}
\newcommand{\rb}{\right )}

\begin{document}

\begin{flushright}
IP-BBSR-2016-7
\end{flushright}

\vspace{7mm}

\begin{center}

{\LARGE \textsc{Hair on non-extremal}} \\

\vspace{6mm}

{\LARGE \textsc{D1-D5  bound states}}

\vspace{15mm}

{\large
\textsc{Pratik Roy${}^{1, 2}$, ~ Yogesh K.~Srivastava${}^{2, 3}$, ~ Amitabh Virmani${}^{1, 2}$
}}

\vspace{13mm}

${}^{1}${Institute of Physics, Sachivalaya Marg, \\ Bhubaneshwar 751005, Odisha, India }\\
\vspace{4mm}
${}^{2}${Homi Bhabha National Institute, Training School Complex, \\ Anushakti Nagar, Mumbai 400085, India}\\
\vspace{4mm}
${}^{3}${National Institute of Science Education and Research (NISER), \\ Bhubaneswar, P.O. Jatni, Khurda 752050, Odisha, India}

\vspace{18mm}

\textsc{Abstract}
\vspace{0.3 cm}

\end{center}
{\small
\noindent
We consider a truncation of type IIB supergravity on four-torus where in addition to the Ramond-Ramond 2-form field, the Ramond-Ramond axion $(w)$ and the NS-NS 2-form field $(B)$ are also retained. In the  $(w, B)$ sector we construct a linearised perturbation  carrying only left moving momentum on two-charge non-extremal D1-D5 geometries of Jejjala, Madden, Ross and Titchener. The perturbation is found to be smooth everywhere and normalisable. It is constructed by matching to leading 
order solutions of the perturbation equations in the inner and outer regions of the geometry.
}

\thispagestyle{empty}

\newpage

\baselineskip=13.5pt
\parskip=3pt

\tableofcontents

\section{Introduction}

Certain earlier studies in black hole physics attempted to find hair on black holes, hoping that if black holes admit large number of hair parameters then they would be analogous to excited atoms. This analogy, if viable, could in turn provide some handle on understanding microscopic origin of black hole entropy. In such studies one looks for smooth and normalisable solutions of perturbations over a black hole background. Those attempts, however, could not find black hole hair, and the results led to the so-called no-hair hypothesis. A version of it states that regardless of the matter model, the end point of gravitational collapse is characterised by conserved charges only. 
Evidence in favour of the hypothesis is presented in the form of  no-hair theorems; evidence against the hypothesis is presented by explicit construction of hairy black holes. Typically the counter-examples violate some energy conditions, or have non-minimal couplings, or have non-canonical kinetic terms. See, e.g.,~references \cite{Bekenstein:1996pn, Herdeiro:2015waa, Cardoso:2016ryw}  for reviews on these developments.   The absence of such hair modes suggested that  the analogy between black holes and atoms is not a viable one.\footnote{These statements are under
the assumption that the test fields inherit all spacetime symmetries. Dropping this assumption, recent studies \cite{Herdeiro:2015waa, Cardoso:2016ryw} have shown that 
stationary black holes do admit hair where one can draw a parallel with states of the hydrogen atom.}

String theory  has had numerous successes in black hole physics including some key results on the
statistical mechanical interpretation of the Bekenstein-Hawking entropy. In string theory at weak couplings it has been shown in great detail that certain intersecting D-brane systems with the same charge and mass as the black hole have precisely the requisite number of states \cite{Strominger:1996sh, Mandal:2010cj}.  These calculations have been successfully extended in numerous ways, though several issues still remain unanswered. Notably, gravity (or bulk) description of these microstates is not yet fully understood. The fuzzball proposal \cite{Lunin:2001jy, Mathur:2005zp, Mathur:2012zp} addresses this question.  It argues that the black hole event horizon and its interior are replaced with quantum bound states with size of the order of the event horizon. In this program large classes of smooth supergravity solutions have been constructed, and their weak coupling description understood. The evidence from the construction of  such black hole microstates suggests that the true quantum bound state indeed has a size of order the event horizon. This proposal completely changes the traditional picture of the black hole horizon and its interior.

In the early days of the fuzzball proposal, perturbative hair modes\footnote{The term ``hair'' has different meanings in the black hole microstate literature. Depending on the context sometimes it has been used to refer to `neck' degrees of freedom and sometimes to refer to `cap' degrees of freedom. In this paper, the hair mode we construct has its support in the neck region and it also adds momentum charge to the background geometry.} were used to exhibit the first examples of microstates for the three-charge system.
Mathur, Saxena, and Srivastava (MSS) \cite{Mathur:2003hj} pioneered these investigations. They constructed a  BPS perturbation over certain smooth BPS D1-D5 geometries. The perturbation was found to be smooth everywhere and normalisable. This situation needs to be contrasted with black holes. As discussed above black holes seem not to admit hair, but black hole microstates in the sense of the fuzzball proposal do admit such hair. In fact such hair modes lead to further microstates, as the hair modes smoothly deform the geometry in a controlled manner to another geometry. In some cases non-linear completion of the hair modes can also be worked out, where it has been shown that the linearisation of the non-linear solution leads to the hair mode. Over the years, this circle of ideas has been a recurrent theme in the fuzzball literature \cite{Lunin:2004uu, Mathur:2011gz, Lunin:2012gp, Mathur:2012tj, Shigemori:2013lta, Martinec:2015pfa,Martinec:2014gka} and together with other developments has culminated in the construction and understanding of the BPS superstratum \cite{Bena:2014qxa, Bena:2015bea, Giusto:2013bda, Giusto:2015dfa, Bena:2016agb} --- supergravity solutions parameterised by arbitrary functions of two variables.

The key aim of the present paper is to demonstrate that certain non-supersymmetric fuzzballs also admit hair modes. We show this by adapting MSS study to a non-supersymmetric setting. Let us recall that on the CFT side MSS started with a particularly simple 2-charge state: the single unit spectral flow (both left and right) of the NS-NS vacuum $ |0\rangle_\rom{NS}$. To add `hair' to this state, they considered the action of a chiral primary on $ |0\rangle_\rom{NS}$ to get $ |\psi\rangle_\rom{NS}$. Then they considered a descendant of this state in the chiral primary multiplet obtained by the action of the left lowering operator of angular momentum \be J^-_{0}{}_L |\psi\rangle_\rom{NS}. \label{state} \ee Under one unit of  spectral flow on both the left and the right sectors on the state  \eqref{state} we get an  RR state  that carries one unit of linear momentum. This final state is thought of as hair on the 2-charge state in the gravity description.

On the supergravity side, the geometry of the starting 2-charge state has an asymptotically flat region connected to spectral flowed AdS$_3 \ \times $ S$^3$ core via the throat region \cite{Maldacena:2000dr, Balasubramanian:2000rt}.
 MSS constructed a supergravity perturbation on top of this geometry in a matched asymptotic expansion in the following steps:
 \begin{enumerate}
\item The perturbation solution in the inner region was obtained  by  applying spectral flows on the appropriate solution in the chiral primary multiplet on top of  global AdS$_3 \ \times $ S$^3$.
\item In the outer region the perturbation solution was obtained by explicitly solving the equations of motion and choosing the normalisable solution. 
\item As the last step they matched the inner and the outer region solutions. It is very non-trivial that such a matching can be done. 
\end{enumerate}

In this paper we retrace the above steps without much additional technical complications for the non-extremal 2-charge D1-D5 geometries of Jejjala, Madden, Ross and Titchener (JMaRT) \cite{Jejjala:2005yu}. 

Unlike the motivation of MSS,  our aim in this work is not to show that the non-extremal D1-D5-P geometries exist perturbatively;  classes of such non-linear solutions are already 
known \cite{Bossard:2014yta, Bossard:2014ola, Bena:2015drs}. 
Our aim in this work is to emphasise that perturbative techniques used for finding supersymmetric  fuzzballs can be taken over to non-supersymmetric set-ups as well. We adapt the MSS scheme to leading  order only (because of technical complications) in the matched asymptotic expansion for the JMaRT set-up. Extending the computation to higher orders is significantly more difficult and is not attempted in this paper. We leave that for future work, though certain ground work for this is set up in appendix \ref{app:harmonics}. In the BPS case, it is now understood that MSS perturbation is the linearisation of a non-linear solution of IIB supergravity \cite{Giusto:2013rxa}. Perhaps it may  be feasible to find the non-linear version of the perturbative solution constructed in this paper. 

The rest of the paper is organised as follows. We start with a brief review of the 2-charge non-extremal D1-D5 bound states of  Jejjala, Madden, Ross and Titchener \cite{Jejjala:2005yu} in section \ref{sec:2charge}.  In section \ref{sec:perturbation} linearised perturbation equations are solved to leading order. We end with a brief discussion in section \ref{sec:discussion}. In order to carry out our analysis numerous properties of  harmonics on S$^3$ are required. Relevant properties are collected in appendix \ref{app:harmonics} together with various conventions and notation we follow in the main text.

\section{Two charge non-extremal D1-D5 bound states}

\label{sec:2charge}
In this section we briefly review the 2-charge non-extremal D1-D5 bound states of  Jejjala, Madden, Ross and Titchener (JMaRT) \cite{Jejjala:2005yu}. We first present the supergravity solutions, illustrate  how to take limits to the inner and outer regions, and then briefly mention their CFT interpretation. The JMaRT solutions have received renewed attention in the last  couple of years \cite{
Gimon:2007ps, Katsimpouri:2014ara, deLange:2015gca, Chakrabarty:2015foa, Chakrabarty:2016nbu}.

 The self-dual two-charge JMaRT solution  is obtained by setting $\delta_1  = \delta_5$ in the 2-charge solutions of \cite{Jejjala:2005yu}.  In order to connect to the notation of \cite{Mathur:2003hj} we also replace 
 \be
a_1 \to - a, \qquad \theta \to \frac{\pi}{2} -\theta,  \qquad  \phi \leftrightarrow \psi.
 \ee
The six-dimensional metric then simplifies to
\begin{align}
d s^2 \  = \ \ & \frac{1}{H} \big{[} - (f-M) (d t + (f-M)^{-1} M c^2 a \sin^2\theta d \phi)^2 + f (d y - f^{-1} M s^2 a \cos^2 \theta d\psi)^2 \big{]}  \nn  \\
	    & +  H\lb \frac{d r^2}{r^2+a^2-M} + d\theta^2   + \frac{r^2}{f}\cos^2\theta d\psi^2 + \frac{r^2+a^2-M}{f-M}\sin^2\theta d\phi^2 \rb, \label{JMaRT_metric}
\end{align}
where $ s= \sinh \delta, c = \cosh \delta, f=r^2+a^2\cos^2\theta$, and $H=f+M s^2$.

The RR two-form supporting this metric takes the form
\begin{eqnarray}
  C  &=& - \frac{M a s c}{ H} \sin^2\theta \   d y\wedge d\phi - \frac{Ma  s c}{H} \cos^2\theta \ dt\wedge d\psi  
  \nn \\ & &
  -\frac{M s c }{H}d t\wedge d y - \frac{Ms c }{H}(r^2+Ms ^2) \sin^2\theta d\phi\wedge d\psi. \label{JMaRT_C2}
\end{eqnarray}
The resulting field strength $F = dC$  can be easily checked to be self-dual $F= \star F$. Moreover,  $F \wedge \star F = 0$, i.e.,
\be
F_{ABC} F^{ABC} = 0.
\ee
Due to the self-duality property of $F$ it immediately follows that $d \star F =0$.
It can also be easily checked that the six-dimensional (simplified) Einstein equations
\be
R_{AB} - \frac{1}{4} F_{A CD} F_{B}{}^{CD} =0,
\ee
 are satisfied. The gauge charges in five-dimensions upon reduction along $y$ are simply 
\be
Q_1 = Q_5 = Q = M s c, \qquad Q_p = 0.
\ee

 The determinant of the $(y,\theta, \phi, \psi)$ part of the metric \eqref{JMaRT_metric}  vanishes  at $r=0$. This signals that a circle direction shrinks to zero size there. The Killing vector with closed orbit that has zero norm at $r=0$ is
\be
\xi = \partial_y + \frac{a}{ M s^2} \partial_\psi.
\ee
That is, the direction that goes to zero size at $r=0$ is $y$ at fixed $(\tilde \psi, \phi)$ where
\be
\tilde \psi = \psi -  \frac{a}{ M s^2} y.
\ee
In order to make $y \to y  + 2 \pi R$ a closed orbit  at fixed $(\tilde \psi, \phi)$ we require 
\be
\frac{a R}{M s^2} = m \in \mathbb{Z}. \label{integer}
\ee
To have smooth degeneration of the $y$ circle near $r=0$ we choose the radius $R$ as
\be
R = \frac{M s^2}{\sqrt{a^2 - M}}. \label{size_R}
\ee
Inserting $R$ from \eqref{size_R} into the integer quantisation  condition \eqref{integer}  fixes $M$ to be
\be
M=a^2\lb 1-\invert{m^2}\rb. \label{integer_mass}
\ee

Upon taking $m=1$ we recover the supersymmetric solutions of \cite{Maldacena:2000dr, Balasubramanian:2000rt} as follows. Setting $m=1$ in \eqref{integer_mass} we see that $M =0$. In order to keep charges  $ Q_1 = Q_5 = Q =  M s c$  finite we must take $\delta \to \infty$. In this limit  metric \eqref{JMaRT_metric} further simplifies to 
\begin{align}
d s^2 \  = \ \ & \frac{1}{H} \bigg{[} - f \lb d t +  \frac{a Q}{f} \sin^2\theta d \phi \rb^2 + f \lb d y - \frac{a Q}{f} \cos^2 \theta d\psi \rb^2 \bigg{]}  \nn  \\
	    & +  H\lb \frac{d r^2}{r^2+a^2} + d\theta^2   + \frac{r^2}{f}\cos^2\theta d\psi^2 + \frac{r^2+a^2}{f}\sin^2\theta d\phi^2 \rb \label{MSS_metric},
\end{align}
with $H  = f + Q$.
This metric can be rewritten in the following more recognisable form
\bea{
d s^2 = \ &-\invert{h}(d t^2-d y^2)+hf\left(d\theta^2+\frac{d r^2}{r^2+a^2}\right)-\frac{2aQ}{hf}\left(\cos^2\theta d yd\psi+\sin^2\theta d td\phi\right)\nn\\
	     &  +h\left[\left(r^2+\frac{a^2Q^2\cos^2\theta}{h^2f^2}\right)\cos^2\theta d\psi^2+\left(r^2+a^2-\frac{a^2Q^2\sin^2\theta}{h^2f^2}\right)\sin^2\theta d\phi^2\right],
}
where 
\be
h ~=~ \frac{H}{f} ~ =  ~ 1+\frac{Q}{f}.
\ee
This form of the 2-charge metric was the starting point in \cite{Mathur:2003hj}.
Upon setting $m=1$ the RR 2-form field \eqref{JMaRT_C2} simplifies to
\begin{eqnarray}
  C  &=&-  \frac{Q a \sin^2\theta}{Q + f} d y\wedge d\phi - \frac{ Q a\cos^2\theta}{Q + f}  dt\wedge d\psi  
  \nn \\ & &
  -\frac{Q }{Q+f}d t\wedge d y - \frac{Q}{Q+f}(r^2+Q) \sin^2\theta d\phi\wedge d\psi,
\end{eqnarray}
which also matches with the corresponding expression in  \cite{Mathur:2003hj} upon making a simple gauge transformation $C \to C + d \Lambda$ with $\Lambda =  Q \phi d\psi$.

\subsection*{Inner and outer regions}
In the next section we solve certain perturbation equations in a matched asymptotic expansion.  In order to do this we split the JMaRT geometry in two overlapping regions: the inner and outer regions. We  solve the perturbations equations in two regions separately and match the solutions in the overlap region.

\subsubsection*{The `Inner' Region}

To describe the above two-charge configuration from the  AdS/CFT perspective we must take a near-decoupling limit where
a large inner region involving AdS$_3 \times$ S$^3$ develops. This inner region is  coupled to flat space via a neck region.
The decoupling limit is achieved by taking the large $R$ limit, i.e., by taking 
\be
\epsilon = \frac{\sqrt{Q}}{R}  \ll 1,\label{epsilon}
\ee
while keeping the charge $Q$ and other parameters of the geometry fixed. 
In this limit the region $ r \ll \sqrt{Q}$ is identified as the inner region. This amounts to taking $H \approx Q$ in \eqref{JMaRT_metric} and taking $M s c \approx Q$ in the cross terms in the metric. From \eqref{JMaRT_metric} and integer quantisation \eqref{integer} we get an asymptotically  AdS$_3 \times$ S$^3$ metric,
\begin{align}
d s^2 \  = \ \ & \frac{1}{Q} \big{[} - (f-M) (d t + (f-M)^{-1} a Q \sin^2\theta d \phi)^2 + f (d y - f^{-1} a Q \cos^2 \theta d\psi)^2 \big{]}  \nn  \\
	    & +  Q\lb \frac{d r^2}{r^2+a^2-M} + d\theta^2   + \frac{r^2}{f}\cos^2\theta d\psi^2 + \frac{r^2+a^2-M}{f-M}\sin^2\theta d\phi^2 \rb, \label{JMaRT_AdS}
\end{align}
supported by the $C$-field
\begin{align}
C_{ty} &= \frac{r^2}{Q},  & C_{\psi \phi} &= Q \sin^2 \theta \\
C_{y \phi} &= - a \sin^2 \theta, & C_{\psi t } &= a \cos^2\theta.
\end{align}
The parameter $M$ in equation \eqref{JMaRT_metric}  can be replaced in favour of the integer $m$ and rotation parameter $a$ through equation \eqref{integer}.

Under the change of coordinates
\be
\psi_\rom{NS}=\psi-\frac{m}{R}y, ~ \ \ \phi_\rom{NS}=\phi-\frac{m}{R}t,	\label{spectral}
\ee
metric  \eqref{JMaRT_AdS} becomes
\beeq
d s^2 & = & -\frac{1}{Q}\left(r^2+\frac{a^2}{m^2}\right)d t^2+\frac{r^2}{Q}d y^2+ Q\left(r^2+\frac{a^2}{m^2}\right)^{-1}d r^2 \nn \\ 
& & +  \ Q(d\theta^2+\cos^2\theta d\psi_\rom{NS}^2+\sin^2\theta d\phi_\rom{NS}^2). \label{AdS_S3}
\eeeq
Setting $m=1$ the above metric can be recognised as exactly the AdS$_3 \times$ S$^3$ metric used by MSS in the NS-NS sector.  The coordinate transformation \eqref{spectral}
is the gravity dual of the D1-D5 CFT spectral flow transformation. Defining 
\be
a' = \frac{a}{m}
\ee
metric \eqref{AdS_S3} can be written in a more convenient form
\be
d s^2  =  -\frac{(r^2 + a'{}^2)}{Q}d t^2+\frac{r^2}{Q}d y^2+ \frac{Q}{(r^2 + a'{}^2)} d r^2 +  \ Q(d\theta^2+\cos^2\theta d\psi_\rom{NS}^2+\sin^2\theta d\phi_\rom{NS}^2).
\label{inner_region_aprime}
\ee
The $F$-field supporting the inner region metric \eqref{inner_region_aprime} is simply
\beeq
F_{tyr} &=& \frac{2r}{Q},   \label{inner_region_F1} \\
F_{\theta \psi \phi} &=& 2 Q \sin \theta \cos \theta. \label{inner_region_F2}
\eeeq

From equations \eqref{integer} and \eqref{epsilon} we also have the relation
\be
\epsilon = \frac{a}{m \sqrt{Q}}.
\ee
As a result, the inner region limit $r \ll \sqrt{Q}$ together with $\epsilon \ll 1$ can operationally be thought of as an expansion in powers of $\frac{r}{\sqrt{Q}}, \ \frac{a}{m \sqrt{Q}}$.
In the NS-NS coordinates via an explicit calculation we see that the metric expansion is in powers of $\epsilon^2$. Since in this paper we are only interested in perturbation to $\mathcal{O}(\epsilon^0)$ we will not be concerned with the corrections to the inner region metric.

\subsubsection*{The `Outer' Region}

The outer region 
$ 
\epsilon \sqrt{Q} \ll r<\infty
$
 has the metric 
\be
d s^2=-\frac{r^2}{Q+r^2}(d t^2-d y^2)+\frac{(Q+r^2)}{r^2}d r^2+(Q+r^2)(d\theta^2+\cos^2\theta d\p^2+\sin^2\theta d\phi^2), 	\label{metricouter}
\ee
to the leading order. 
The 2-form $C$-field supporting the outer region metric is simply
\beeq
C_{ty} = - \frac{Q}{Q +r^2} &\implies& F_{tyr} = \frac{2 Q r}{(Q+r^2)^2}\label{outer_C_1}, \\
C_{\psi \phi} =  Q \sin^2 \theta &\implies & F_{\theta \psi \phi} = 2 Q \sin \theta \cos \theta. \label{outer_C_2}
\eeeq
The overlap of the inner and outer regions is in the region
\be
\epsilon \sqrt{Q} \ll r \ll \sqrt{Q}.
\ee
The outer region limit can be operationally thought of as a power series expansion in 
$\frac{a}{m r}, \frac{a}{m\sqrt{Q}}$.

The outer region metric \eqref{metricouter} does not have any non-extremal features. In particular, it does not look like non-extremal D1-D5 black hole. This is because in the JMaRT construction the fuzzball  and  the black hole parameter spaces are mutually exclusive.

\subsection*{CFT description}
The 2-charge JMaRT solution corresponds to $m = 2 s + 1$ units of left and $m = 2 s+ 1$ units of right spectral flows on the NS-NS vacuum. Since equal amount of left and right spectral flows are involved \be h=\bar h, \ee and the state has no net linear momentum. When $m=1$, i.e., $s = 0$, we get a supersymmetric configuration, a Ramond-Ramond ground state. For other values of $s$, supersymmetry is broken as both the left and right sectors carry (equal amount of)  excitations. CFT description of JMaRT solution has been explored in great detail in the literature \cite{Jejjala:2005yu,  Chakrabarty:2015foa}, we refer the reader to these references for further details.

\section{Perturbation}
\label{sec:perturbation} 
In this section we construct the RR axion and NS-NS 2-form field perturbation to leading order in both the inner and the outer regions and show their matching. 
The linearised perturbation equations are,
 \be
 H_{ABC}+\invert{3!}\epsilon_{ABCDEF}H^{DEF}+wF_{ABC}=0, 				\label{hstarh}
 \ee
 \be
 \Box w-\invert{3}F^{ABC}H_{ABC}=0,								\label{boxw}
 \ee
where $F$ is simply $dC$, 
\be F_{ABC} = 3\partial_{[A}C_{BC]} = \partial_A C_{BC} + \partial_B C_{CA} + \partial_C C_{AB}. 
\ee
These equations are precisely the equations studied by Mathur, Saxena, and Srivastava \cite{Mathur:2003hj}.  The main difference\footnote{Our $F$ and $H$ notation conforms to more standard string theory notation, but it is interchanged compared to reference \cite{Mathur:2003hj}.} is that in \cite{Mathur:2003hj}, the background considered is supersymmetric, whereas in our study it is not.  An embedding of these equations in IIB theory was studied in \cite{Giusto:2013rxa} for BPS solutions.\footnote{In earlier versions of this paper on the arXiv (also in the published version) we attempted to embed these equations in IIB theory using the truncation 
studied in \cite{Duff:1998cr}, but later realised that our analysis was not fully correct. In the present version we have removed that appendix. The interpretation of $(w, B)$ as the RR axion and IIB NS-NS field is motivated by the results of \cite{Giusto:2013rxa}. We thank the anonymous referee for alluding this problem to us. We believe that the embedding discovered in   \cite{Giusto:2013rxa} is applicable to our set-up as well, though we do not have a complete argument as of now.}

\subsection{CFT description}
\label{CFT_description}
Before getting in to the supergravity analysis, it is useful to understand the CFT state to which the perturbation we construct below corresponds to. In the NS-NS sector we act with a chiral primary operator with $h=l, \bar h = l$ on the NS-NS vacuum. Let us denote the resulting state by $|\psi\rangle_\rom{NS}$. On this state we act with angular momentum lowering generators both on the left and on the right  $k$ and $\bar k$ times respectively:
\be
(J_0^-{}_{L})^k (J_0^-{}_{R})^{\bar k} \ket{\psi}_\rom{NS}.
\ee
Next, on the resulting state we act with $m= 2 s + 1$ (odd) units of spectral flows both on the left and on the right. As a result we get an excited state in the R-R sector 
\be
\left[ (J_0^-{}_{L})^k (J_0^-{}_{R})^{\bar k}  |\psi\rangle\right]_\rom{R}^{(2s+1)}. 
 \ee
For general values of the $l, k, \bar k, m = 2s +1$ the state thus obtained carries both left and right moving excitations.  As we explain momentarily, we tune $\bar k$ such that `for the perturbation' $\bar h$ is zero, but $h\neq 0$. Such a perturbation does not carry any `extra energy' and is tractable to leading order in our supergravity analysis.

\subsection{Perturbation at leading order}
We first construct the inner region perturbation and then the outer region.
\subsubsection{Inner region}
In the inner region \eqref{inner_region_aprime}--\eqref{inner_region_F2} the perturbation equation \eqref{hstarh} gives,
\be
H_{tyr}+\frac{r}{Q^2}\left(2wQ+\frac{H_{\theta\psi\phi}}{\sin\theta\cos\theta}\right)=0, \label{inner_main_1}
\ee
when all three indices are taken either on  AdS$_3$ or on S$^3$.  When one index is on AdS$_3$ and the others on S$^3$ we get the same equations as when one index is on S$^3$ and the others on AdS$_3$. We get
\be
H_{\m\n a}+\invert{6}\epsilon_{\m\n a\r bc}g^{\r\r'}g^{bb'}g^{cc'}H_{\r'b'c'}=0, \label{inner_main_2}
\ee
where the Greek indices refer to AdS coordinates and the lower case Latin indices to S$^3$ coordinates. We now expand the various fields in harmonics on S$^3$ as (see appendix \ref{app:decompose})
\begin{align}
&B_{\m\n} =  b_{\m\n}^{I_1}Y^{I_1}, \label{expansion1} \\
& B_{ab}  =  b^{I_1}\tilde{\epsilon}_{ab}{}^{c}\del_cY^{I_1},  \\
& B_{\m a} =  b_{\m}^{I_3}Y_{a}^{I_3}, \\
& w =  w^{I_1}Y^{I_1}, \label{expansion4}
\end{align}
where $\tilde{\epsilon}_{abc}$ is the epsilon-tensor on the \emph{unit} 3-sphere. We use the convention $\tilde{\epsilon}_{\theta \psi \phi} = + \sin \theta \cos \theta$.   To avoid notational clutter we do not write explicit summation signs with spherical harmonics; sum over appropriate indices is understood.
We can use various properties of these harmonics. Spherical harmonics on S$^3$ are appropriate representations of $\mathfrak{so}(4) \simeq \mathfrak{su}(2) \times \mathfrak{su}(2)$. For scalar harmonics $Y^{(l,l)}_{(m,m')}$ the upper labels tell the representations of $ \mathfrak{su}(2)$'s and the corresponding lower index tells the state in the representation. Vector harmonics come in two variety: one class with $\mathfrak{su}(2) \times \mathfrak{su}(2)$ representations $(l,l+1)$ and the other $(l+1,l)$. We use the following properties of these harmonics
\be
\grad^2_{S^3}Y^{I_1}=-C(I_1)Y^{I_1}, \qquad \del_aY_b^{I_3}-\del_bY_a^{I_3}=\eta(I_3)\tilde{\epsilon}^c{}_{ab}Y_c^{I_3},
\ee
where $C(I_1)=4l(l+1)$ for $(l,l)$ representation and 
$\eta(I_3) = - 2(l+1)$ for $I_3 = (l+1,l)$ and  $\eta(I_3) = + 2(l+1)$ for $I_3 = (l,l+1)$.

Upon substituting the expansions \eqref{expansion1}--\eqref{expansion4} in  \eqref{inner_main_1}-- \eqref{inner_main_2} and using properties of spherical harmonics we find the following simplified equation
\be
\grad_\rom{AdS}^2b^{I_1}+\left(2w^{I_1}-\frac{C(I_1)}{Q}b^{I_1}\right)=0,			\label{boxbi}
\ee
where $\grad_\rom{AdS}^2$ refers to the scalar Laplacian with respect to the three-dimensional `global' AdS metric
\be
d s^2_\rom{AdS}  =  -\frac{(r^2 + a'{}^2)}{Q}d t^2+\frac{r^2}{Q}d y^2+ \frac{Q}{(r^2 + a'{}^2)} d r^2.
\label{metricAdS3}
\ee
In arriving at \eqref{boxbi} we also use the convention $\epsilon_{tyr} = + \sqrt{-g_\rom{AdS}}$.

Similar manipulations on equation (\ref{boxw}) gives,
\be
Q\grad_\rom{AdS}^2w^{I_1}-C(I_1)w^{I_1}-8\left(w^{I_1}-\frac{C(I_1)}{Q}b^{I_1}\right)=0.			\label{boxwi}
\ee
We now have a system of differential equations
\be
\grad^2 \begin{pmatrix} b^{I_1} \\ w^{I_1} \end{pmatrix} = \begin{pmatrix} C(I_1)/Q & -2 \\  -8C(I_1)/Q^2 &  (C(I_1)+8)/Q^2 \end{pmatrix} \begin{pmatrix} b^{I_1} \\ w^{I_1} \end{pmatrix}.
\ee
We can diagonalize the system. The problem then becomes that of solving the equation
\be
\grad_\rom{AdS}^2f-m^2f=0.
\ee
Since $t$ and $y$ are Killing directions of \eqref{metricAdS3}, we can take the solution to be of the form
\be
f(t,y,r)=e^{-i\omega t+ik_y y}h(r),
\ee
and solve for $h(r)$. The solution corresponding to a chiral primary 
 in the inner region is \cite{Maldacena:1998bw, Mathur:2003hj}
\beeq
&& w = \frac{e^{-2ilt(a'/Q)}}{Q(r^2+a'{}^2)^l}\hat{Y}_\rom{NS}^{(l)},\\
&& B_{ab} = B \tilde \epsilon_{abc}\del^c\hat{Y}_\rom{NS}^{(l)}, \\
&& B_{\m\n} =  \invert{\sqrt{Q}}\epsilon_{\m\n\l}\del^{\l}B\hat{Y}_\rom{NS}^{(l)},
\eeeq
where
\beeq
&& B = \frac{e^{-2ilt(a'/Q)}}{4l(r^2+a'{}^2)^l},  \\
&& \hat{Y}_\rom{NS}^{(l)}=(Y_{(l,l)}^{(l,l)})_\rom{NS}=\sqrt{\frac{2l+1}{2}}\frac{e^{-2il\phi_\rom{NS}}}{\pi}\sin^{2l}\theta.
\eeeq
In these formulae all the quantities related to the sphere are with respect to unit sphere. Quantities with Greek indices refer to metric \eqref{metricAdS3}.
This perturbation has quantum numbers
\begin{align}
h^\rom{NS}&= l, & m^\rom{NS} &= l, & \bar h^\rom{NS} &= l, & \bar m^\rom{NS} &= l.  \label{QM1}
\end{align}

As mentioned in the beginning of this section, we consider lowering this state with $J_0^-{}_{L}$ and $J_0^-{}_{R}$ $k$ and $\bar k$ times respectively, i.e.,
\be
(J_0^-{}_{L})^k (J_0^-{}_{R})^{\bar k} \ket{\psi}_\rom{NS}. \label{NS_state}
\ee
Such states have quantum numbers
\begin{align}
h^\rom{NS}&= l, & m^\rom{NS} &= l-k, & \bar h^\rom{NS} &= l, & \bar m^\rom{NS} &= l-\bar k.  \label{QM2}
\end{align}
In the NS-NS sector we can immediately write down the supergravity fields for the state \eqref{NS_state}:
\beeq
&& w = \left( \frac{e^{-2ilt(a'/Q)}}{Q(r^2+a'{}^2)^l} \right) \mathcal{Y}_{\rom{NS}}   		,\\
&& B_{ab} = B \tilde \epsilon_{abc}\del^c \mathcal{Y}_{\rom{NS}}   		, \\
&& B_{\m\n} =  \left( \invert{\sqrt{Q}}\epsilon_{\m\n\l}\del^{\l}B \right) \mathcal{Y}_{\rom{NS}}   		,
\eeeq
where
\beeq 
B &=& \frac{e^{-2ilt(a'/Q)}}{4l(r^2+a'{}^2)^l},  \\
\mathcal{Y}_\rom{NS} &=& Y^{(l, l)}_{(l-k, l - \bar k)}= 
\underbrace{
\left\{\pounds_{\xi^-_L} \pounds_{\xi^-_L} \cdots \pounds_{\xi^-_L} \right\}}_\text{$ k-$times}  
\underbrace{
\left\{\pounds_{\xi^-_R} \pounds_{\xi^-_R} \cdots \pounds_{\xi^-_R} \right\} }_\text{$\bar k-$times} 
\hat Y^{(l)}_\rom{NS} (\theta, \phi_{\rom{NS}}   , \psi_{\rom{NS}}  ).
\eeeq

Under a spectral flow  on the left-moving sector with parameter $\alpha$ the quantum numbers change as:
\beeq
h' &=& h - \alpha m + \alpha^2 \frac{c}{24}, \\ 
m' &=& m -\alpha \frac{c}{12}.
\eeeq
Spectral flow of the background (NS-NS vacuum) gives 
\beeq
h_\rom{BG}' &=&  \alpha^2 \frac{c}{24},  \\
m_\rom{BG}' &=&  -\alpha \frac{c}{12}.
\eeeq
Therefore for the perturbation alone, we identify
\beeq
h_\rom{P}' &=& h - \alpha m, \\
 m_\rom{P}' &=&  m,
\eeeq
so that the full $h' = h_\rom{BG}' + h_\rom{P}'.$

Applying $m$ (odd) units of spectral flow on the left and right sectors on the perturbation \eqref{QM2}, we get a perturbation in the Ramond-Ramond sector with quantum numbers
\begin{align}
h^\rom{R}_\rom{P}&= l - m (l - k), & m^\rom{R}_\rom{P} &= l-k, & \bar h^\rom{R}_\rom{P} &=  l - m (l - \bar k), & \bar m^\rom{R}_\rom{P} &= l - \bar k.  \label{QM3}
\end{align}
From these expressions we see that in general $\omega = h + \bar h$ is not equal to $|k_y|$ where $k_y = h - \bar h$. In such a situation, not all the energy of the perturbation is tied to the S$^1$ momentum. The residual energy goes in the `radial motion'.  For technical convenience we work with perturbations where all energy is tied to the S$^1$ momentum, i.e., we demand
 \be 
\bar h^\rom{R}_\rom{P} = 0.
\ee This fixes $\bar k$ in terms of the integers $l$ and $m$, 
\be
\bar k= \frac{l}{m}(m-1).
\ee
The requirement that $\bar k $ must be an integer between $1$ and $2 l$ constrains the values of $l$ and $m$.

The perturbation in the R-R coordinates then looks like
\begin{align}
& w	\		= \	 \ 	\invert{Q}\frac{\mathcal{E}}{(r^2+a'{}^2)^l} \mathcal{Y}_{\rom{R}},   		\label{winner}		\\
& B_{\theta\p} \	= \ \ 	\invert{4l}\frac{\mathcal{E}}{(r^2+a'{}^2)^l}\cot\theta~\del_{\phi} \mathcal{Y}_{\rom{R}},			\\
& B_{\theta\phi} \	= \	 \ 	-\invert{4l}\frac{\mathcal{E}}{(r^2+a'{}^2)^l}\tan\theta~\del_{\psi} \mathcal{Y}_{\rom{R}},		\\
& B_{\psi\phi} \	= \ \ 	\invert{4l}\frac{\mathcal{E}}{(r^2+a'{}^2)^l}\sin\theta\cos\theta~\del_{\theta} \mathcal{Y}_{\rom{R}},	\\
& B_{ty} \		= \ \	-\invert{2Q^2}\frac{r^2\mathcal{E}}{(r^2+a'{}^2)^l} \mathcal{Y}_{\rom{R}},				\\
& B_{yr} \		= \ \	\frac{i a'}{2Q}\frac{r\mathcal{E}}{(r^2+a'{}^2)^{l+1}} \mathcal{Y}_{\rom{R}},
\end{align}
together with the following terms generated via the spectral flow \eqref{spectral},
\begin{align}
& B_{t\theta} \	= \ - \frac{m a'}{Q} B_{\phi\theta} \ =  \ 	-\frac{m a'}{Q}\frac{1}{4l}\frac{\mathcal{E}}{(r^2+a'{}^2)^l}\tan\theta~\del_{\psi} \mathcal{Y}_{\rom{R}}, \\
& B_{t\psi} \	= \ - \frac{m a'}{Q} B_{\phi\psi}  \ =  \	\frac{m a'}{Q}\frac{1}{4l}\frac{\mathcal{E}}{(r^2+a'{}^2)^l}\sin\theta\cos\theta~\del_{\theta} \mathcal{Y}_{\rom{R}} ,\\
& B_{y\theta} \	= \ - \frac{m a'}{Q} B_{\phi\theta} \ =  \ 	\frac{m a'}{Q}\frac{1}{4l}\frac{\mathcal{E}}{(r^2+a'{}^2)^l}\cot\theta~\del_{\phi} \mathcal{Y}_{\rom{R}},		\\
& B_{y\phi} \	= \ - \frac{m a'}{Q} B_{\psi\phi}  \ = \	-\frac{m a'}{Q}\frac{1}{4l}\frac{\mathcal{E}}{(r^2+a'{}^2)^l}\sin\theta\cos\theta~\del_{\theta} \mathcal{Y}_{\rom{R}},	
\end{align}
where 
\beeq
\mathcal{E} &=& \exp\left[-\frac{i}{R}\left( l - m (l - k) \right) u  \right].
\eeeq
In writing these expressions we have defined $u=t +y$, and we have used the spectral flow transformations \eqref{spectral} to write angular dependence in the R-R coordinates $(\phi, \psi)$.

Two comments are in order here: \begin{enumerate}
 \item From these equations we see that the perturbation carries $l - m (l - k)$ units of linear momentum. Since $k$ is an integer taking values between $0$ and $2l$, we have in fact constructed $2l+1$ hair modes. All these hair modes belong to the same chiral primary multiplet. 
 
 \item One might be concerned that the left conformal weight $h_\rom{P}^\rom{R}$ in \eqref{QM3} is negative when $m> 1$. This is not a problem, as we note that the total $h$ values are never negative. One can think that the perturbation \emph{draws} energy from the filled Fermi sea, which makes the $h_\rom{P}^\rom{R}$ value negative. 
 
\end{enumerate}

\subsubsection{Outer region}
We now want to solve the equations (\ref{hstarh}), (\ref{boxw}) in the outer region \eqref{metricouter} -- \eqref{outer_C_2}. We are interested in a solution that falls off at inifinity in a normalisable manner and matches smoothly with the inner region solution. We proceed in exactly the same manner as in the inner region. Substituting spherical harmonic expansion \eqref{expansion1}--\eqref{expansion4} in equation (\ref{hstarh}) we get 
\be
\frac{1}{r}\left( - \partial_t^2 + \partial_y^2 \right) b^I_1 + \partial_r\left(\frac{r^3}{(Q+r^2)^2} \partial_r b^{I_1}\right)
+\frac{r}{(Q+r^2)^2}\left[2 Q w^{I_1} - C(I_1) b^{I_1} \right]=0. \label{outer_main_1}
\ee
Similar manipulations on equation (\ref{boxw}) gives,
\beeq
&&\frac{Q+r^2}{r^2}\left( - \partial_t^2 + \partial_y^2 \right) w^{I_1} + \frac{1}{r(Q+r^2)}\partial_r\left(r^3\partial_r w^{I_1}\right)
\\ && 
- \frac{C(I_1)}{Q+r^2} w^{I_1} 
- \frac{8Q}{(Q+r^2)^3}\left[Q w^{I_1} - C(I_1) b^{I_1} \right]=0, \label{outer_main_2}
\eeeq
For the perturbation with the $(t,y)$ dependence of the form of  
\be
\mathcal{E} =\exp\left[-\frac{i}{R}\left( l - m (l - k) \right) u  \right],
\ee
the $\left( - \partial_t^2 + \partial_y^2 \right)$ combination is zero. Therefore the above equations simplify to
\beeq
&& \partial_r\left(\frac{r^3}{(Q+r^2)^2} \partial_r b^{I_1}\right)
+\frac{r}{(Q+r^2)^2}\left[2 Q w^{I_1} - C(I_1) b^{I_1} \right] =0, \label{outer_main_1_1}\\
&& \frac{1}{r(Q+r^2)}\partial_r\left(r^3\partial_r w^{I_1}\right)
- \frac{C(I_1)}{Q+r^2} w^{I_1} 
- \frac{8Q}{(Q+r^2)^3}\left[Q w^{I_1} - C(I_1) b^{I_1} \right]=0. \label{outer_main_1_2}
\eeeq

These equations can be integrated as follows.  We  write equations \eqref{outer_main_1_1}  and \eqref{outer_main_1_2} as 
\beeq
&& \frac{d^2 \bs}{dr^2} + P_1 \frac{d \bs}{dr} + Q_1 \bs = R_1,  \label{outer_main_2_1} \\  
&& \frac{d^2 \ws}{dr^2} + P_2 \frac{d \ws}{dr} + Q_2 \ws = R_2, \label{outer_main_2_2}
\eeeq
where 
\begin{align}
P_1 &= \frac{3 Q -r^2}{r(Q+r^2)},  & 
P_2 &= \frac{3}{r}, &
Q_1 &= \frac{-C(I_1)}{r^2}, & \\
Q_2 &= \frac{-C(I_1)}{r^2} -\frac{8Q^2}{r^2(Q+r^2)^2},&
 R_1 &= \frac{-2 Q \ws}{r^2},&
 R_2 &= \frac{-8Q C(I_1)\bs}{r^2(Q+r^2)^2}.&
\end{align}
We now eliminate first derivative terms from equations \eqref{outer_main_2_1}  and \eqref{outer_main_2_2}. We write $\bs = u_1 v_1$ and $\ws = u_2 v_2$ and choose $u_1 ,u_2$ such that 
first derivative terms cancel out. We get 
\be
u_1 = 
\frac{Q+r^2}{r^\frac{3}{2}} \ \ , \ \ 
u_2 = 
\frac{1}{r^\frac{3}{2}}
\ee
and equations for $v_1 , v_2$ become
\be
\frac{d^2 v_1}{dr^2} +  A_1 v_1 = -2 A_3 v_2   \ \ ,  \ \  \frac{d^2 v_2}{dr^2} + A_2 v_2 = -8 C(I_1) A_3 v_1
\ee
where 
\be
A_1 = \frac{8Q}{(Q+r^2)^2} -\frac{3+4 C(I_1)}{4 r^2}, \ A_2 = -\frac{8Q^2}{r^2(Q+r^2)^2} -\frac{3+4 C(I_1)}{4 r^2} ,  \ A_3 = 
\frac{Q}{r^2 (Q+r^2)}.
\ee
Writing this in the matrix form we have 
\be
\frac{d^2}{dr^2} \left(\begin{array}{c}v_1\\ v_2 \end{array}\right) +  \left(
  \begin{array}{cc} 
    A_1 & 2 A_3  \\ 
    8 C(I_1) A_3 & A_2  \\ 
  \end{array}
  \right)\left(\begin{array}{c}v_1\\ v_2 \end{array}\right) =0.
\ee 
Diagonalising this matrix and solving the second order uncoupled ordinary differential equations for normalisable solutions,  we get 
\beeq
\bs  &=& \frac{1}{4l} \frac{ 1}{r^{2l} }, \\
\ws  &=&   \frac{ 1}{ r^{2l}(r^2+Q)}.
\eeeq

The perturbation solution decaying at infinity therefore takes the form
\beeq
&& w=\frac{\mathcal{E}}{(Q+r^2)r^{2l}}  \mathcal{Y}_{\rom{R}},			\\					
&& B_{ab}=B\tilde \epsilon_{abc}\del^c  \mathcal{Y}_{\rom{R}},  \\
&& B_{\m\n}=\invert{\sqrt{Q+r^2}}\e_{\m\n\rho}\del^{\rho}B~ \mathcal{Y}_{\rom{R}},  \\
&& B=\invert{4l}\frac{\mathcal{E}}{r^{2l}}.
\eeeq
Here all quantities related to the sphere are with respect to the unit sphere. Writing out all the components explicitly, the solution looks as follows:
\begin{align}
& w	\		= \ \	\frac{\mathcal{E}}{(Q+r^2)r^{2l}} \mathcal{Y}_{\rom{R}},			\label{wouter}			\\
& B_{\theta\p} \	= \ \	\invert{4l}\frac{\mathcal{E}}{r^{2l}}\cot\theta~\del_{\phi} \mathcal{Y}_{\rom{R}}				,			\\
& B_{\theta\phi} \	= \ \	-\invert{4l}\frac{\mathcal{E}}{r^{2l}}\tan\theta~\del_{\psi} \mathcal{Y}_{\rom{R}}			,				\\
& B_{\p\phi} \		= \ \	\invert{4l}\frac{\mathcal{E}}{r^{2l}}\sin\theta\cos\theta~\del_{\theta} \mathcal{Y}_{\rom{R}}				,		\\
& B_{ty} \ 		= \ \	-\invert{2(Q+r^2)^2}\frac{\mathcal{E}}{r^{2l-2}} \mathcal{Y}_{\rom{R}}				,			\\
& B_{tr} \		= \ \	- \invert{4l} \frac{1}{r^{2l+1}}(\partial_y \mathcal{E}) \mathcal{Y}_{\rom{R}}				,					\\
& B_{yr} \		= \ \ - \invert{4l} \frac{1}{r^{2l+1}}(\partial_t \mathcal{E}) \mathcal{Y}_{\rom{R}}			.\end{align}

\subsubsection{Matching}

We now want to see that the solutions that we have found for the inner and outer regions join onto each other smoothly in the overlap of the two regions, 
\be
\epsilon \sqrt{Q} \ll r \ll \sqrt{Q}.
\ee 
We can choose to do the matching at any value of $r$ in the above range. We choose the matching radial coordinate to be the geometric mean of the two ends
\be
r \sim  \sqrt{\epsilon}\sqrt{ Q}.
\ee
In this region, the scalar $w$ in both the inner (\ref{winner}) and the outer (\ref{wouter}) regions has the same leading order behaviour,
\be
w \simeq \frac{\mathcal{E}  \mathcal{Y}_{\rom{R}}}{Qr^{2l}}.
\ee
We see that the scalar does indeed match.  To compare the NS-NS 2-form, we can construct the field strength $H=d B$ and compare the behaviour in the two regions in a given orthonormal frame. More simply we notice that from the inner region point of view 
\be
\frac{1}{(r^2 + a'{}^2)} \approx \frac{1}{r^{2}},
\ee
and from the the outer region point of view 
\be
\frac{1}{(Q + r^2)} \approx \frac{1}{Q},
\ee
near the matching surface. We find that components $B_{\theta \psi}, B_{\theta \phi}, B_{\psi \phi}, B_{ty}$ match readily. The components that do not match $B_{tr}, B_{yr}$ and all components generated by spectral flow in the inner region, namely $B_{t\theta}, B_{t\psi}, B_{y\theta}, B_{t\phi}$ are seen to be higher order terms. These terms will agree once inner region and outer region calculations are corrected by higher order terms.

\section{Discussion}
\label{sec:discussion}
In this paper we have constructed `hair' on non-extremal  two-charge bound states of Jejjala et al \cite{Jejjala:2005yu}. Our perturbation is `extremal' as it is of the form $e^{i c u}$. To the best of our knowledge this is the first example of a hair mode on a non-extremal smooth geometry, and also the first example beyond JMaRT of a non-extremal fuzzball with identified CFT interpretation. We have constructed the perturbation by a matching procedure.  In the inner region our background is simply $m$ units of both left and right spectral flows on the NS-NS vacuum. To construct our perturbation we have taken an appropriate state in the chiral primary multiplet on top of the NS-NS vacuum  (global AdS$_3 \times$ S$^3$) and have applied left and right spectral flows  to get a perturbation over the R-R state of interest. Such a perturbation in our linearised analysis can be matched on to a normalisable solution in the outer region. We have carried out the matching procedure to the leading order. We find it quite remarkable that we are able to add such a hair on the non-extremal D1-D5 bound states. 

By altering the above construction slightly one can construct another class of hair modes as follows. Instead of lowering $|\psi\rangle_\rom{NS}$ with $J_{L,R}^-$ one can first apply even units (say, $2p$)  of spectral flows on $|\psi\rangle_\rom{NS}$ to get an excited state in the NS-NS sector $|\psi\rangle_\rom{NS}^{(2p)}$. 
 Then one can apply $J_{L,R}^-$ as above, and finally apply $ 2(m-p)+1$ units of spectral flows to get the background state corresponding to the 2-charge JMaRT. By such a procedure one will get different values for $h_\rom{P}^\rom{R}$ and $\bar h_\rom{P}^\rom{R}$. By demanding $\bar h_\rom{P}^\rom{R} =0$ one can construct a different class of hair, which can also be matched to a decaying solution in the outer region. Other variations on this procedure are also possible. It can be interesting to work out these solutions explicitly.

The JMaRT solutions are well known to suffer from ergoregion instability \cite{Cardoso:2005gj}. The reader may wonder how have we managed to add hair on an unstable geometry. The answer lies in the fact that we work in a sector completely different from the sector in which the instability of the geometry is analysed in \cite{Cardoso:2005gj}. We added \emph{extremal} perturbation in the Ramond-Ramond axion and NS-NS two-form sector. The full haired solution is most likely unstable to scalar perturbations. In fact one can easily convince oneself that it is not possible to add hair modes on the JMaRT solution in the minimally coupled scalar sector \cite{Mathur:2003hj}. It is not clear to us if one can add \emph{non-extremal} perturbation to the JMaRT solutions in the $(w, B)$ sector, i.e., perturbations with both $h_\rom{P}^\rom{R}$ and $\bar h_\rom{P}^\rom{R}$ non-zero.

There are several ways in which our study can be extended. It will be interesting to perform the analysis at higher order in $\epsilon$ like it was done in \cite{Mathur:2003hj}. This is likely to be technically intricate. There are two main sources of complications: $(i)$ the spherical harmonic expansions are significantly more involved compared to the MSS study, and $(ii)$ features related to non-extremality of the background will appear in the perturbation analysis at higher orders.  We hope to report on these calculations at some point in the future. As mentioned in the introduction, in the BPS case, it is now understood that the MSS perturbation is the linearisation of a non-linear solution of IIB supergravity \cite{Giusto:2013rxa}. Perhaps it may  be feasible to find the non-linear version of the perturbative solution constructed in this paper.

\subsection*{Acknowledgment}
This project grew out of discussions between AV and YS at  the 2015 IISER Bhopal SERC Preparatory School. We thank the organisers for their kind hospitality. AV and PR  thank the organisers of the GIAN Lecture Course on Black-hole Information Loss Paradox at IIT Gandhinagar for providing a stimulating atmosphere where this project was finalised. We are grateful to Samir Mathur and David Turton for discussions.  We also thank David Turton and Bidisha Chakrabarty for reading through a draft of the manuscript. The work of AV is supported in part by the DST-Max Planck Partner Group ``Quantum Black Holes'' between IOP, Bhubaneshwar and AEI, Golm.

\appendix
\section{Spherical harmonics on S$^3$}
\label{app:harmonics}
In this appendix we list relevant expressions and properties of spherical harmonics we needed. For explicit expressions we follow the conventions of \cite{Mathur:2003hj}, but there are typos in their expressions that we fix below.
The metric on the unit 3-sphere is
\be
ds^2=d\q^2+\sin^2\q d\phi^2+\cos^2\q d\psi^2. \label{unit_sphere}
\ee
The scalar and vector harmonics satisfy the orthonormality conditions
\begin{align}
\int d\Omega (Y^{I_1})^*(Y^{I_1^{\prime}} ) \ &= \ \delta^{I_1,I_1^{\prime} },   \\
\int d\Omega (Y_a^{I_2})^* Y^{I_2^{\prime}a} \ &= \ \delta^{I_2,I^{\prime}_2}.
\end{align}
We use the following notation from \cite{Mathur:2003hj}
\begin{align}
 \hat{Y}^{(l)} \  & \equiv \  Y^{(l,l)}_{(l,l)},	&					\\
 Y^{(l)} \ &\equiv \  Y^{(l,l)}_{(l-1,l)}	,		&			\\
Y^{(l+1)} \ &\equiv \ Y^{(l+1,l+1)}_{(l-1,l)},		&		\\
Y^{(l+1,l)}_a \ &\equiv \ Y^{(l+1,l)}_{a(l-1,l)},	&			\\
Y^{(l,l+1)}_a \ &\equiv \ Y^{(l,l+1)}_{a(l-1,l)},	&			\\
Y^{(l-1,l)}_a \ &\equiv \ Y^{(l-1,l)}_{a(l-1,l)},		&		\\
Y^{(l+2,l+1)}_a \ &\equiv \ Y^{(l+2,l+1)}_{a(l-1,l)}.&		
\end{align}

The above harmonics are,
\begin{align}
\hat Y^{(l)} & \equiv \  Y^{(l,l)}_{(l,l)} = \frac{\sqrt{2l+1}}{\sqrt{2}}\frac{e^{-2il\phi}}{\pi}\sin^{2l}\q						\\
Y^{(l)}&=-\frac{\sqrt{l(2l+1)}}{\pi}e^{-i(2l-1)\phi + i\psi}\sin^{2l-1}\q \cos\q		\\
Y^{(l+1)}&=\frac{\sqrt{(2l+1)(2l+3)}}{2\pi}e^{-i(2l-1)\phi + i\psi}((l-1)+(l+1)\cos2\q)\sin^{2l-1}\q\cos\q
\end{align}
\begin{align}
Y_{\q}^{(l+1,l)} \ &= \ -\frac{e^{-i(2l-1)\phi+i\psi}}{4\pi}\frac{\sin^{2l-2}\q}{\sqrt{l+1}}((2l^2-l+1)+(l-1)(2l+1)\cos2\q)					\\
Y_{\phi}^{(l+1,l)} \ &= \ i\frac{e^{-i(2l-1)\phi+i\psi}}{4\pi}\frac{\sin^{2l-1}\q\cos\q}{\sqrt{l+1}}((2l^2-5l-1)+(2l^2+3l+1)\cos2\q)			\\
Y_{\psi}^{(l+1,l)} \ &= \ -i\frac{e^{-i(2l-1)\phi+i\psi}}{4\pi}\frac{\sin^{2l-1}\q\cos\q}{\sqrt{l+1}}((2l^2+3l-1)+(l+1)(2l+1)\cos2\q)
\end{align}
\begin{align}
Y^{(l,l+1)}_{\q} \ & = \ -\frac{e^{-i(2l-1)\phi+i\psi}}{4\pi}\sqrt{\frac{4l(2l+1)}{l+1}}\sin^{2l-2}\q((l-1)+l\cos2\q)				\\
Y^{(l,l+1)}_{\phi} \ & = \ i\frac{e^{-i(2l-1)\phi+i\psi}}{4\pi}\sqrt{\frac{4l(2l+1)}{l+1}}\sin^{2l-1}\q\cos\q((l-2)+(l+1)\cos2\q)		\\
Y^{(l,l+1)}_{\psi} \ & = \ i\frac{e^{-i(2l-1)\phi+i\psi}}{4\pi}\sqrt{\frac{4l(2l+1)}{l+1}}\sin^{2l-1}\q\cos\q(l+(l+1)\cos2\q)
\end{align}
\begin{align}
Y^{(l-1,l)}_{\q} \ &= \ \frac{e^{-i(2l-1)\phi+i\psi}}{2\pi}\sqrt{2l-1}\sin^{2l-2}\q					\\
Y^{(l-1,l)}_{\phi} \ &= \ -i\frac{e^{-i(2l-1)\phi+i\psi}}{2\pi}\sqrt{2l-1}\sin^{2l-1}\q\cos\q			\\
Y^{(l-1,l)}_{\psi} \ &= \ -i\frac{e^{-i(2l-1)\phi+i\psi}}{2\pi}\sqrt{2l-1}\sin^{2l-1}\q\cos\q
\end{align}
\begin{align}
Y^{(l+2,l+1)}_{\q} \ =& \ -\frac{e^{-i(2l-1)\phi+i\psi}}{8\pi} \sqrt{\frac{3}{l+2}}\sin^{2l-2}\q \left[ (l-1)(2l^2+l+1) \right.				\nn		\\
				\ &   \  \left.+\frac{2(4l^3-l+3)\cos2\q}{3}+\frac{(l-1)(l+1)(2l+3)\cos4\q}{3} \right]										\\
Y^{(l+2,l+1)}_{\phi} \ =& \ i\frac{e^{-i(2l-1)\phi+i\psi}}{4\pi} \sqrt{\frac{3}{l+2}}\sin^{2l-1}\q \cos\q \left[\frac{2l^3-3l^2+3l+4}{2} \right.  \nn 		\\
				\ 	& \left. \invert{3}(4l^3-13l-9)\cos2\q + \frac{(l+1)(l+2)(2l+3)}{6}\cos4\q \right]										\\
Y^{(l+2,l+1)}_{\psi} \ =& \ -i\frac{e^{-i(2l-1)\phi+i\psi}}{4\pi} \sqrt{\frac{3}{l+2}}\sin^{2l-1}\q \cos\q \left[\frac{l(2l^2+5l-1)}{2}  \right. 	\nn  		\\
				\ 	& \left. +\invert{3}(l+1)(4l^2+8l-3)\cos2\q+\frac{(l+1)(l+2)(2l+3)}{6}\cos4\q  \right].
\end{align}

\subsection*{Raising and lowering the harmonics}
The Killing forms for the metric \eqref{unit_sphere} can be written as, see e.g., appendix A of \cite{Kanitscheider:2006zf}
\begin{align}
\xi_{1L} \ &= \   \cos (\psi +\phi )d\theta+\sin \theta  \cos \theta   \sin (\psi +\phi ) d(\psi -\phi )  ,			\\
\xi_{2L} \ &= \ -\sin (\psi +\phi )d\theta   + \sin \theta  \cos \theta  \cos (\psi +\phi ) d(\psi -\phi ),  		\\
\xi_{3L} \ &= \ \cos ^2\theta d\psi +\sin ^2\theta  d\phi 									,	\\
\xi_{1R} \ &= \ - \sin (\psi -\phi )d\theta + \sin \theta  \cos \theta  \cos (\psi -\phi )d(\psi +\phi )	,	\\
\xi_{2R} \ &= \  \cos (\psi -\phi )d\theta +\sin \theta  \cos \theta   \sin (\psi -\phi )d(\psi +\phi )	,	\\
\xi_{3R} \ &= \ - \cos ^2\theta  d\psi + \sin ^2\theta  d\phi.
\end{align}
From these expressions, we can readily obtain the six Killing vectors on S$^3$ (using the order $\theta, \phi, \psi$),
\begin{align}
\xi_{1L}^{\mu} \ &= \ (\cos(\phi+\psi),-\cot\theta\sin(\phi+\psi),\tan\q\sin(\phi+\psi)),		\\
\xi_{2L}^{\mu} \ &= \ (-\sin(\phi+\psi),-\cot\q\cos(\phi+\psi),\tan\q\cos(\phi+\psi)),					\\
\xi_{3L}^{\mu} \ &= \ (0,1,1)								,					\\
\xi_{1R}^{\mu} \ &= \ (\sin(\phi-\psi),\cot\q\cos(\phi-\psi),\tan\q\cos(\phi-\psi))	,				\\
\xi_{2R}^{\mu} \ &= \ (\cos(\phi-\psi),-\cot\q\sin(\phi-\psi),-\tan\q\sin(\phi-\psi))	,				\\
\xi_{3R}^{\mu} \ &= \ (0,1,-1).
 \end{align}
These vectors generate the $SU(2)\times SU(2)$ algebra as
\be
[\xi_{aL},\xi_{bL}]=2\e^{abc}\xi_{cL}, \qquad  [\xi_{aR},\xi_{bR}]=2\e^{abc}\xi_{cR},  \qquad   [\xi_{aL},\xi_{bR}]=0,
\ee
 where $a,b,c={1,2,3},$ and $\e$ is the completely antisymmetric symbol with $\e^{123}=1.$
 We can now use these Killing vectors to define the raising and lowering operators on the harmonics. Defining
\begin{align}
\xi_L^+ &= \xi_{1L}+i\xi_{2L}, \qquad    \qquad  \xi_L^- = \xi_{1L}-i\xi_{2L},   \\
\xi_R^+ &= \xi_{2R}-i\xi_{1R},   \qquad \qquad  \xi_R^- =  \xi_{2R}+i\xi_{1R}, \label{raising_lowering}
\end{align}
one can check that the harmonics satisfy the following relations (here $\pounds_{\xi}$ denotes the Lie derivative along the vector $\xi$),
\begin{align}
\frac{i}{2} \pounds_{\xi_{3L}} \hat{Y}^{(l)} \ &= \ + l \hat{Y}^{(l)}, \\
\frac{i}{2} \pounds_{\xi_{3R}} \hat{Y}^{(l)} \ &= \ + l \hat{Y}^{(l)}, \\
 \pounds_{\xi^+_L} \hat{Y}^{(l)} \ &= \ 0, \\
\pounds_{\xi^+_R} \hat{Y}^{(l)} \ &= \ 0,
\end{align}
confirming that  $\hat{Y}^{(l)}$'s indeed correspond to highest weight states. Moreover, one can check that
\begin{align}
\pounds_{\xi^-_L}\hat{Y}^{(l)}  \ &= \ -2\sqrt{2l}Y^{(l)},									\\
\pounds_{\xi^+_L}Y^{(l)} \ &= \ 2\sqrt{2l}\hat{Y}^{(l)}	,								\\
\pounds_{\xi^-_L}(\pounds_{\xi^-_R}Y^{(l+1,l+1)}_{(l,l+1)}) \ &= \ - 8\sqrt{(l+1)(2l+1)}Y^{(l+1)},	\\	
\pounds_{\xi^-_R}(\pounds_{\xi^-_L} Y^{(l,l+1)}_{a(l,l+1)}) \ &= \ -8\sqrt{l(l+1)}Y^{(l,l+1)}_a,		\\
\pounds_{\xi^+_R}(\pounds_{\xi^+_L}Y_a^{(l+2,l+1)}) \ &= \ 8\sqrt{3}(l+1)Y^{(l+2,l+1)}_{a(l,l+1)}.
\end{align}
The pre-factors in these equations are convention dependent, e.g., they depend on the normalisation in definition \eqref{raising_lowering}. Using these properties we were able to fix typos in expressions of vectors harmonics in \cite{Mathur:2003hj}.

Other properties of spherical harmonics we need are
\beeq
\nabla^2 Y^{I_1} &=& - 4 l (l+1)  Y^{I_1}, \\
\nabla_{[a}\nabla_{b]}Y^{I_1} &=& 0,\\
\nabla^a Y_a^{I_1} & = &0.
\eeeq
The vector harmonics fall into two classes of $SU(2) \times SU(2)$ representations: $(l, l+1)$ or $(l+1,l)$. We have
\be
\nabla_a Y_b^{I_1} - \nabla_b Y_a^{I_1} = \zeta(I_1) \epsilon_{abc} Y^{I_1 c} \label{curl},
\ee
where
$
\zeta(I_1) = + 2 (l+1)
$
for the
$(l, l+1)$
representations
and
$
\zeta(I_1) = - 2 (l+1)
$
for the
$(l+1,l)$
representations. We use the convention $\epsilon_{\theta \psi \phi} = + \sin \theta \cos \theta$ \cite{Mathur:2003hj}.

\subsection*{Decomposition}
\label{app:decompose}
For completeness let us  recall (see e.g., \cite{Gerlach:1978gy, Ishibashi:2004wx}) that a vector field on S$^3$ can be uniquely decomposed into its longitudinal and transverse parts,
\be
V_a = \nabla_a S + S_a,
\ee
where $S$ is a scalar and $S_a$ is a vector transverse  on S$^3$,
$\nabla^a S_a = 0. $ The scalar field $S$ and the transverse vector $S_a$ can be Fourier decomposed on the sphere in terms of spherical harmonics.

Since a two-form is dual to a vector on S$^3$, it also admits a decomposition in terms of scalar and vector spherical harmonics. Let the 2-form be $B_{ab}$
\be
B_{ab} = \epsilon_{abc} V^c. \label{dualB}
\ee
Now note that on the one hand,
\be
\epsilon_{abc} \nabla^b (\epsilon_{mn}{}^{c} B^{mn}) = - 2 \nabla^b B_{ba},
\ee
and on the other hand from \eqref{dualB},
\be
\epsilon_{abc} \nabla^b (\epsilon_{mn}{}^{c} B^{mn}) = \epsilon_{abc}  \nabla^{b} V^{c} = \epsilon_{abc}  \nabla^{b} S^{c} .
\ee
For non-trivial vector harmonics the quantity $\epsilon_{abc}  \nabla^{b} S^{c}$ is non-zero, cf.~\eqref{curl}. Therefore, in de-Donder gauge when
\be
\nabla^b B_{ba} =0,
\ee
the decomposition of an arbitrary 2-form $B_{ab}$ is solely in terms of derivatives of scalar harmonics \cite{Deger:1998nm, Maldacena:1998bw, Mathur:2003hj},
\be
B_{ab} = \sum_{I_1} b^{I_1}\epsilon_{ab}{}^{c}\partial_c Y^{I_1}.
\ee

\end{document}